\documentclass[aps,prd,preprintnumbers,showpacs]{revtex4}
\usepackage[dvips]{graphicx}
\begin{document}

%
%

\preprint{Nisho-1-2009}
\title{Chiral Anomaly and Decay of Color Electric Field}
\author{Aiichi Iwazaki}
\address{International Economics and Politics, Nishogakusha University,\\ 
Ohi Kashiwa Chiba  277-8585, Japan.}   
\date{April 8, 2009}
\begin{abstract}
Using the formula of chiral anomaly, 
we discuss the pair production of quarks under color electric field $\vec{E}$ 
without addressing explicit formula of quark's wavefunctions. 
The production is assumed to occur under the effect of color magnetic field $\vec{B}$ as well as the color electric field.
We obtain the life time $t_c$ of the color electric field 
in the limit of
$B\gg E$. 
Applying it to the glasma in high-energy heavy-ion collisions,
we find that $t_c\simeq 10Q_s^{-1}$ with saturation momentum $Q_s$.
\end{abstract}
\hspace*{0.3cm}
\pacs{12.38.-t, 24.85.+p, 12.38.Mh, 25.75.-q  \\
Schwinger mechanism, Chiral Anomaly, Color Glass Condensate}
\hspace*{1cm}

\maketitle

Initial states of color gauge fields (glasma) produced in high-energy heavy-ion collisions
have recently received much attention. The gauge fields are longitudinal color electric and magnetic fields.
It is expected that
the decay of the glasma leads to thermalized quark gluon plasma (QGP). 
Although the color electric field generated in the collisions has been previously discussed\cite{iwazaki} from
a phenomenological point of view,  
the presence of the classical gauge fields 
is recently argued on the basis of a fairly reliable
effective theory of QCD at high energy, that is, a model of color glass condensate (CGC)\cite{colorglass}.
 
The glasma is homogeneous in the longitudinal
direction and inhomogeneous in the transverse directions. Hence,
they form color electric and magnetic flux tubes extending in the longitudinal direction. 
In the previous papers\cite{iwa,itakura,hii} we have shown a possibility that 
the famous Nielsen-Olesen instability\cite{nielsen} makes the color magnetic field
unstable. The possibility has been partially confirmed by
the comparison of our results with those of numerical simulations\cite{venugopalan,berges}. 
As is well known, Nielsen-Olesen unstable modes 
inhomogenuos in the longitudinal direction arise 
under the homogeneous color magnetic field. 
On the other hand,
the simulations
show that
the glasma become unstable when small fluctuations of gauge fields inhomogeneous in the
longitudinal direction are present. This instability in the glasma has been identified with the Nielsen-Olesen instability
in our analyses.

In this paper we discuss the decay of the color electric field
under the effect of the color magnetic field.
The decay is caused by the pair production ( Schwinger mechanism\cite{schwinger} ) of quarks and antiquarks,
and by the acceleration of these particles.
For instance, the quarks are accelerated by the color electric field after their production,
so that the electric field loses its energy.
Although the pair production under both electric and 
magnetic fields has been discussed in previous papers\cite{tanji}(see the references therein),
the decay of the electric field 
under the effect of the magnetic field have been only discussed in a paper\cite{tanji}. 
The distinctive point in our paper is to use chiral anomaly 
for analysing the pair production without addressing the explicit forms of quark's wavefunctions.
Using the anomaly 
we can exactly obtain the production rates of the quarks because the anomaly implicitly involves
the quantum effects of the pair production. 
However, the use is limited only 
for the system with massless charged fermions under
the presence of both electric and magnetic fields.

Before discussing the pair production of quarks,
we first consider the pair production of electrons and positrons under electric and magnetic fields.
We assume that they are massless and  
that the fields $\vec{B}$ and $\vec{E}$ are spatially homogeneous and parallel (or antiparallel) with each other.
They are oriented in the $z$ direction.
Then, the energies of electrons with charge $-e<0$ and positrons with charge $e>0$ 
under the magnetic field $B=|\vec{B}|$ are given by

\begin{equation}
\label{1}
E_N=\sqrt{p_z^2+2NeB} \quad \mbox{(parallel)} \quad \mbox{and} \quad E_N=\sqrt{p_z^2+eB(2N+1)} \quad \mbox{(antiparallel)},
\end{equation}
where $p_z$ denotes momentum in the $z$ direction and 
integer $N\ge 0$ does Landau level. The term of
``parallel" (``antiparallel") implies magnetic moment parallel (antiparallel) to $\vec{B}$. 
The magnetic moment of electrons (positrons) is antiparallel (parallel) to their spin.
Thus,
electrons (positrons) with spin antiparallel (parallel) to $\vec{B}$ can have zero energy states
in the lowest Landau level; their energy spectrum is given by $E_{N=0}=|p_z|\ge 0$. 
On the other hand, the other states cannot be zero energy states; their energy spectra are given
by $\sqrt{p_z^2+eBM}\ge eBM\ge eB$ with positive integer $M$.
In a sense they are ``massive" states whose masses increase with $B$.  

When the color electric field is imposed,
the pair production of electrons and positrons with the energies $E_N$
occurs.
However, it is not probable that any states are produced 
with an equal production rate. Indeed, high energy states are
hard to be produced while low energy states are easy to be produced.
Thus,
the ``massive" states are hard to be produced by weak electric field $E \ll B$.
In particular, they cannot be produced in the limit of $B \to \infty $.
Only products in the limit are the pairs of electrons and positrons in zero energy states ( $E_{N=0}=|p_z|=0$ ).

Note that owing to the fermi statistics
only fermions with $p_z=0$ are produced,
since the other states with $p_z\neq 0$ have already been occupied.
After the production of the fermions with $p_z=0$, their momenta
increase with time, $p_z(t)=\pm e\int^t dt' E$, owing to the acceleration by the electric field.
Hence we obtain the momentum distribution $\tilde{n}(p_z)\propto \theta(p_F(t)-p_z)\theta(p_z)$ 
for the particles with positive charge $e>0$ and $\tilde{n}(p_z)\propto \theta(p_F(t)+p_z)\theta(-p_z)$
for the particles with negative charge $-e<0$. 
The fermi momentum $p_F$ is given by $p_F(t)=e\int_0^tdt'E(t')$.
Here we have assumed that the electric field is parallel to $\vec{B}$
and is switched on at $t=0$.  
The momentum distributions agree with those shown in the previous paper\cite{tanji}.
(When $\vec{E}$ is antiparallel to $\vec{B}$, the momentum of the particles 
increases in the different direction from the case of $\vec{E}$ being parallel $\vec{B}$.)

We note that electrons move to the direction antiparallel to $\vec{E}$ while
positrons move to the direction parallel to $\vec{E}$.
Therefore, both electrons and positrons produced by the electric field have right handed helicity
when $\vec{E}$ being parallel $\vec{B}$, while they have left handed helicity when $\vec{E}$ being antiparallel $\vec{B}$. 
(Right or left handed helicity means the state with momentum parallel or antiparallel to spin, respectively.)


Now, we discuss the pair production rate and the decay of the electric field.
The key point is to use the equation of chiral anomaly,

\begin{equation}
\label{chiral}
 \partial_t (n_R-n_L)=\frac{e^2}{4\pi^2}E(t)B
\end{equation}
where $n_R$ ( $n_L$ ) denotes number density of right ( left ) handed chiral fermions; 
$n_{R,L}=\langle\bar{\Psi}\gamma_0(1\pm \gamma_5)\Psi\rangle/2$
in which the expectation value is taken by using a state of electrons and positrons produced
under the electric field. Here, we have assumed spatial uniformness of the chiral current $\vec{j}_5$, that is, $\rm{div}\vec{j}_5=0$.

In accordance with the eq(\ref{chiral}) of the chiral anomaly, 
the rate of chirality change is given by the product of $\vec{E}$ and $\vec{B}$.
It is important to remember that only particles with right ( left ) handed helicity
are produced in the limit of $B\gg E$ when $\vec{E}$ is parallel to $\vec{B}$ ($\vec{E}$ being antiparallel to $\vec{B}$.) 
Since the number density $n$ of electrons
is the same as that of positrons, $n_R=2n$ and $n_L=0$ when $\vec{E}$ being parallel to $\vec{B}$, while $n_R=0$ and $n_L=2n$ 
when $\vec{E}$ being antiparallel to $\vec{B}$.
Thus, we find that the evolution of the number density with time
is determined by the anomaly equation
$2\partial_t n(t)=e^2|E(t)|B/(4\pi^2)$, where the electric field may depend on time.

It is very interesting to see that when $E$ is static, this simple formula of the chiral anomaly reproduces a
result of $n=e^2EB\,t/(4\pi^2)$, which has been 
obtained in the previous paper\cite{tanji}. The agreement of both results
holds only when $B$ is much larger than $E$.
In other words, the previous calculations\cite{tanji} are consistent with the chiral anomaly.

After the production of the particles, the electric field gradually loses its energy 
because it makes the particles accelerate.
The energy of the electric field is transformed into the
energies $\epsilon$ of the particles, so that the energies of the particles increase,

\begin{equation}
\partial_t(\epsilon(t) +\frac{1}{2}E(t)^2)=0
\end{equation} 
where we have neglected the contribution of magnetic field
which would be induced by the electric current of electrons and positrons.
The energy density $\epsilon(t)$ of the particles are given
such as $\epsilon(t)=2n(t)p_F(t)/2=n(t)p_F(t)$.
This is because the momentum distribution $\tilde{n}(p_z)$ ( $n(t)\equiv \pm\int_0^{\pm\infty} dp_z \tilde{n}(p_z)$ )
is given by $\tilde{n}(p_z)=n_0\theta (p_F(t)-|p_z|)\theta(|p_z|)$ as we have discussed.
For example, the energy density of electrons is given such that $\int_{-\infty}^0 dp_z \tilde{n}(p_z)|p_z|=np_F/2$.

Now we have three equations to solve for obtaining the pair production rate $\partial_t n(t)$, etc.,

\begin{equation}
\label{tot}
2\partial_t n(t)=\frac{e^2}{4\pi^2}|E(t)|B, \quad \partial_t(\epsilon(t) +\frac{1}{2}E(t)^2)=0, \quad \mbox{and}
\quad \epsilon(t)=n(t)p_F(t)
\end{equation}
with $p_F(t)=\int_0^t dt'e|E(t')|$.
It is easy to solve the eq(\ref{tot}) with initial conditions $E(t=0)=E_0>0$ and $n(t=0)=0$ by
assuming homogeneous strong magnetic field $B$ independent on $t$,

\begin{equation}
\label{result}
E(t)=E_0\cos(\sqrt{\frac{2\alpha eB}{\pi}}\,t) \quad \mbox{and} \quad 
n(t)=\frac{\alpha E_0B|\sin(\sqrt{\frac{2\alpha eB}{\pi}}\,t)|}{\pi \sqrt{\frac{2\alpha eB}{\pi}}}
\end{equation}
with $\alpha=e^2/4\pi$.

The formula holds rigorously in the limit of $B\gg E_0$ since the particles with masses on the order of $B$ are
suppressed in the production.
The oscillating features of $E(t)$ and $n(t)$ have agreed with those obtained in the previous papers\cite{tanji},
in which the wavefunctions of electrons under the effects of the electric and magnetic fields
have been explicitly used.

From the eq(\ref{result}) we find that the life time of the electric field defined as $E(t_c)=0$
is given by $t_c=\frac{\pi}{2}(\sqrt{\frac{2\alpha eB}{\pi}})^{-1}$.
This implies that the life time
becomes shorter as the magnetic field becomes stronger.
This is because the number of degenerate states per unit area with the energy $|p_z|$
is given by $ eB/2\pi$. Hence, as $B$ becomes larger, the number of
zero energy states becomes larger so that the number density of the produced particles
increases more. In this way, the increase of $B$ causes the short life time of the electric field.

\vspace*{0.2cm}
In Fig.\,1 we have shown the behaviors $E(t)=E_0\cos(\pi \,t/2t_c)$ and $n(t)=\frac{2E_0\alpha Bt_c}{\pi^2}|\sin(\pi\,t/2t_c)|$ 
with $E_0=1$, $t_c=1$ and $\alpha\simeq 1/137$.
When the electric field is switched on at $t=0$, the pair production begins to occur, so that the number density 
of electrons and positrons increases.
Owing to the acceleration of the particles by the electric field, the energy of the electric field gradually decreases and vanishes at $t=t_c$.
At the same time the number density takes the maximum value. Then, the electric field changes its direction
and becomes strong with time. On the other hand, the number density decreases after $t=t_c$. 
This decrease is caused by the pair annihilation of electrons and positrons. Since the direction of the electric field changes
after $t_c$, the direction of the particle acceleration also change.
Consequently, electrons and positrons are moved to overlap with each other so that the pair annihilation
may occur to make the number density decrease. Eventually, it vanishes at $t=2t_c$. Then, the pair creation repeatedly occurs.
Therefore, the oscillation of $E(t)$ and $n(t)$ arises. 

\begin{figure}[t]
\begin{minipage}{.47\textwidth}
\includegraphics[width=6.7cm,clip]{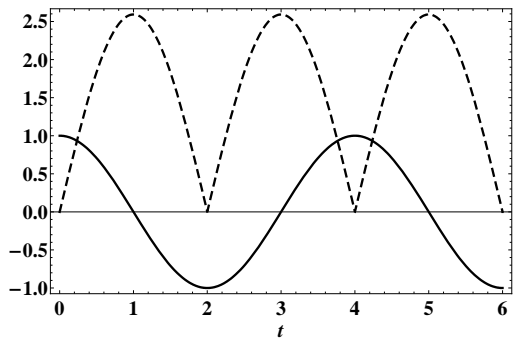}
\label{fig:growth-rate3}
\caption{electric field $E(t)$ (solid line) and number density $n(t)$ (dashed line)}
\end{minipage}
\end{figure} 

We wish to make a comment that although the formula of the chiral anomaly involves all of the quantum effects, 
the cause of the increase or decrease of the number density
is not explicit in the formula. However,
we understand that the increase or decrease of the number density originates with the pair creation or annihilation.
This is because the results shown in this paper have agreed with those obtained by
calculations explicitly including the quantum effects of the pair creation or annihilation.


As a consistency check, we calculate electric current $J$ carried by produced electrons and positrons.
Using a Maxwell equation $\partial_tE(t)=-J(t)$ with no magnetic field,
we obtain 

\begin{equation}
J(t)=E_0\sqrt{\frac{2\alpha eB}{\pi}}\sin(\sqrt{\frac{2\alpha eB}{\pi}}\,t)=2en(t),
\end{equation} 
where we have used the formula $n(t)$ in eq(\ref{result}).
It states that both of electrons with charge density $-en$ and positrons with $en$ 
constitute the electric current $J$ since the velocity of electrons is $-1$, while that of positrons is $+1$
in the unit of light velocity $c=1$. This current is 
conduction current. However, there is no polarization current because the only pair production of the particles with zero energy 
occurs.
This is a very natural result.
Therefore, our results are consistent with Maxwell equations.

\vspace*{0.5cm}
Until now, we have discussed particles production under electric field
using QED. We proceed to discuss the pair production of quarks
in the glasma. We assume that longitudinal color electric and magnetic fields
are oriented in the 3rd direction of SU(3) color space; $E=E^{a=3}$ and $B=B^{a=3}$.
Then, the equation of the chiral anomaly becomes

\begin{equation}
\partial_t (n_R-n_L)=\frac{g^2N_f}{4\pi^2}EB
\end{equation}
where $N_f$ denotes the number of the flavors of massless quarks, i.e. $N_f=2$.
$g$ is the gauge coupling constant. Similarly to the previous case,
only right handed quarks and anti-quarks are produced when $B\gg E$.
Thus, $n_L=0$ and $n_R=2N_f(3-1)n_q=8n_q$ where $n_q$ is the number density of quarks
(or antiquarks) with each flavor and color; they have the same number density $n_q$ as
each other. Since the color electric field is oriented in the 3rd direction of color SU(3) space,
the number of quarks coupled with $E$ are simply $2N_f$.
The coupling strength of the quarks with the gauge field $A_{\mu}^{a=3}$ is given by
$g/2$. Thus, the fermi momentum $p_F$ and the energy density of the quarks
are given by $p_F=\frac{g}{2}\int_0^tE(t')dt'$ and
$\epsilon=4N_fn_qp_F/2=2n_q\int_0^tgE(t')dt'$, respectively. 
(We have assumed that the momentum distributions of the quarks 
are given by $\tilde{n}(p_z)=n_0\theta (p_F(t)-|p_z|)\theta(|p_z|)$.) 
Then, by solving equations for the quarks similar to eq(\ref{tot}),

\begin{equation}
\label{totq}
4\partial_t n_q(t)=\frac{g^2}{4\pi^2}|E(t)|B, \quad \partial_t(\epsilon(t) +\frac{1}{2}E(t)^2)=0, \quad \mbox{and}
\quad \epsilon(t)=4n_q(t)p_F(t).
\end{equation}
we find the evolution of the color electric field and the quark number density with time,

\begin{equation}
\label{result1}
E(t)=E_0\cos(\sqrt{\frac{\alpha_s gB}{\pi}}\,t) \quad \mbox{and} \quad 
n_q(t)=\frac{\alpha_s E_0B|\sin(\sqrt{\frac{\alpha_s gB}{\pi}}\,t)|}{4\pi \sqrt{\frac{\alpha_s gB}{\pi}}},
\end{equation}
with $\alpha_s=g^2/4\pi$.

Therefore, the life time of the color electric field is given by

\begin{equation}
t_c=\frac{\pi}{2}\biggl(\sqrt{\frac{\alpha_s gB}{\pi}}\biggr)^{-1}.
\end{equation}

Although this result holds only in the limit of $B\gg E_0$,
we may use it to approximately estimate the life time of the color electric field
even for $B\sim E_0$. Since the field strength $gE_0$ and $gB$
of the glasma is on the order of $Q_s^2$,
the life time $t_c$ is given by 

\begin{equation}
t_c\simeq 10 \,Q_s^{-1} \simeq 1\,\rm{fm/c} \sim 2\,\rm{fm/c} \quad \mbox{for} \quad 1\,\rm{GeV}\le Q_s \le 2\,\rm{GeV}
\end{equation}
with $\alpha_s=1/4\pi$, namely $g=1$, which is slightly larger than 
the time ($<1$fm/c) required by phenomenological fluid dynamics\cite{hirano} of QGP.

As in the previous case, the color electric field $E(t)$ oscillates with time. 
Phenomenologically, only our concern is the time $t_c$ 
when the color electric field vanishes.
Probably, the produced quark pairs may be thermalized by the time.
Hence, their monemtum distribution takes a different form from $\tilde{n}(p_z)$
assumed in the present paper so that the oscillation would not occur.

\vspace*{0.3cm}
To summarize, using the chiral anomaly we have 
discussed the pair production of quarks under color electric field
without addressing explicit forms of quark's wavefunctions.
The field loses its energy owing to the acceleration of massless quarks produced
by the electric field.
The key point we should stress is that our result 
is exact for the pair production of massless quarks in the limit of $B\gg E$, in which
the contributions of all ``massive" states vanish. 
We have found that the color electric field decay within a time approximately consistent with QGP phenomenology.

\hspace*{1cm}

The author
express thanks to Drs. H. Fujii, N. Tanji of University of Tokyo and Dr. K. Itakura
of KEK for their useful discussion and comments.


\end{document}